\documentclass[a4paper,fleqn,usenatbib]{mnras}
\usepackage[T1]{fontenc}
\usepackage{ae,aecompl}

\usepackage{graphicx}	
\usepackage{amsmath}	
\usepackage{amssymb}
\usepackage{tikz}
\usetikzlibrary{arrows,decorations.pathmorphing}

\title[Chirping Jets from Black Hole Binaries]{Radio Crickets:
  Chirping Jets from Black Hole Binaries Entering their Gravitational
  Wave Inspiral}

\author[Kulkarni and Loeb]{
Girish Kulkarni$^{1}$\thanks{E-mail: kulkarni@ast.cam.ac.uk}
and Abraham Loeb$^{2}$\thanks{E-mail: aloeb@cfa.harvard.edu}
\\
$^{1}$Institute of Astronomy and Kavli
  Institute of Cosmology, University of Cambridge, Madingley Road,
  Cambridge CB3 0HA, UK\\ 
$^{2}$Institute for Theory \& Computation,
  Harvard University, 60 Garden Street, Cambridge, MA 02138, USA\\ 
}

\date{Accepted ---. Received ---; in original form ---}

\pubyear{2015}

\begin{document}
\label{firstpage}
\pagerange{\pageref{firstpage}--\pageref{lastpage}}
\maketitle

\begin{abstract}
  We study a novel electromagnetic signature of supermassive black
  hole binaries whose inspiral starts being dominated by gravitational
  wave (GW) emission.  Recent simulations suggest that the binary's
  member BHs can continue to accrete gas from the circumbinary
  accretion disk in this phase of the binary's evolution, all the way
  until coalescence.  If one of the binary members produces a radio
  jet as a result of accretion, the jet precesses along a biconical
  surface due to the binary's orbital motion.  When the binary enters
  the GW phase of its evolution, the opening angle widens, the jet
  exhibits milliarcsecond scale wiggles, and the conical surface of
  jet precession is twisted due to apparent superluminal motion.  The
  rapidly increasing orbital velocity of the binary gives the jet an
  appearance of a ``chirp.''  This helical chirping morphology of the
  jet can be used to infer the binary parameters.  For binaries with
  mass $10^7$--$10^{10}$ M$_\odot$ at redshifts $z<0.5$, monitoring
  these features in current and archival data will place a lower limit
  on sources that could be detected by eLISA and Pulsar Timing Arrays.
  In the future, microarcsecond interferometry with the Square
  Kilometer Array will increase the potential usefulness of this
  technique.
\end{abstract}

\begin{keywords}
  accretion, accretion discs -- black hole physics -- quasars: supermassive black holes
\end{keywords}

\section{Introduction}
\label{sec:intro}

Binaries of supermassive black holes (BHs) arise naturally as a result
of mergers of galaxies in the context of hierarchical galaxy formation
\citep{2000MNRAS.311..576K, 2002MNRAS.336L..61H, 2008ApJ...676...33D,
  2012MNRAS.422.1306K}.  After a galaxy merger, infalling black holes
lose their angular momentum in three stages before coalescence
\citep{1980Natur.287..307B, 2005LRR.....8....8M, 2011ASL.....4..181C}.
In the first stage, the black holes sink to the center of the
gravitational potential of the merger remnant due to dynamical
friction and form a gravitationally bound binary.  The newly formed
binary then continues to lose its angular momentum and shrink by
scattering ambient gas and stars.  In the final stage of the binary's
evolution, emission of gravitational waves (GWs) becomes the
predominant mode of angular momentum loss, which results in the
coalescence of the black holes.

General relativistic (GR) numerical simulations can predict the
binary's evolution in its final stage up to coalescence
\citep{2005PhRvL..95l1101P, 2006PhRvL..96k1102B, 2006PhRvL..96k1101C}.
This has stimulated interest in predicting related electromagnetic
signals of black hole coalescence.  Effects considered in the
literature include afterglows due to infall of gas on the remnant
black hole \citep{2005ApJ...622L..93M}, electromagnetic variability in
the circumbinary disk due to shocks induced by the sudden mass loss of
the binary following GW emission at coalescence
\citep{2007APS..APR.S1010B, 2008ApJ...676L...5L}, flares from shocked
remnants of accretion disk around the recoiled remnant black hole
\citep{2007PhRvL..99d1103L, 2008ApJ...682..758S, 2009CQGra..26i4032H},
quasi-periodic variability in gas accretion in the early stages of
black hole merger \citep{2008ApJ...672...83M, 2009CQGra..26i4032H,
  2012MNRAS.427.2680K, 2013MNRAS.436.2997D, 2014ApJ...783..134F,
  2015ApJ...807..131S}, evolution in the profiles of broad emission
lines shortly before merger \citep{2013MNRAS.432.1468M},
characteristic evolution in thermal emission due to circumbinary
accretion \citep{2015MNRAS.446L..36F}, brightening of the circumbinary
accretion disk due to viscous dissipation of gravitational waves
\citep{2008PhRvL.101d1101K}, and prompt tidal disruption of ambient
stars by the recoiled remnant \citep{2011MNRAS.412...75S}.

In addition to testing GR in the strong field limit, observations of
electromagnetic counterparts will help localize sources of GWs for
future detectors such as the Evolved Laser Interferometer Space
Antenna (eLISA)\footnote{\url{https://www.elisascience.org}}, which
would be capable of observing the peak GW luminosity for BH
coalescence of $\sim 10^{57}$ erg$/$s out to cosmological distances
\citep{2003CQGra..20S..65H, 2013CQGra..30x4009S}.  Simultaneous
detection of GW and electromagnetic signals from coalescing binaries
will yield rich scientific pay-offs \citep{2003CQGra..20S..65H,
  2005ApJ...629...15H}.  For example, redshift measurement of eLISA
sources could constraint cosmological parameters.  Measurement of the
black hole masses and spins from the GW signal could constrain
accretion physics during coalescence.  But even before eLISA is
launched, a search of electromagnetic signatures may help calibrate
the expected abundance of GW sources.

\begin{figure}
  \begin{center}
    \begin{tikzpicture}
      \draw (0,0) ellipse (3 and 0.5);
      \fill[black] (0,0) circle (1pt);
      \draw [-angle 45,thick] (2,-0.39) -- (3.5,-0.1);
      \draw [-angle 45,thick,dashed,dash pattern = on 4pt off 1pt] (2,-0.39) -- (2,1.5);
      \draw [-angle 45,thick] (2,-0.39) -- (2.5,0.8);
      \draw [-angle 45,thick,red] (2,-0.39) -- (3.5,0.8);
      \fill[black] (-2,0.39) circle (2pt);
      \fill[black] (2,-0.39) circle (2pt);
      \draw (4.1,-0.1) node {\large ${\mathbf{v}}_\mathrm{orb}$};
      \draw (2.7,1.0) node {\large ${\mathbf{v}}_\mathrm{jet}$};
      \draw (4.0,1.0) node {\large $\tilde{\mathbf{v}}_\mathrm{jet}$};
      \draw (2.0,1.8) node {\large $\mathbf{L}_\mathrm{orb}$};
    \end{tikzpicture}
  \end{center}
  \caption{Jet and orbital configuration. The orbital velocity
    $\mathbf{v}_\mathrm{orb}$ causes the jet to precess as its source
    BH moves along its binary orbit. The net jet velocity is
    $\tilde{\mathbf{v}}_\mathrm{jet}={\mathbf{v}}_\mathrm{orb}+{\mathbf{v}}_\mathrm{jet}$,
    where $\mathbf{v}_\mathrm{jet}$ is its ejection velocity of the
    jet material at the source.}
  \label{fig:BH_orbit}
\end{figure}
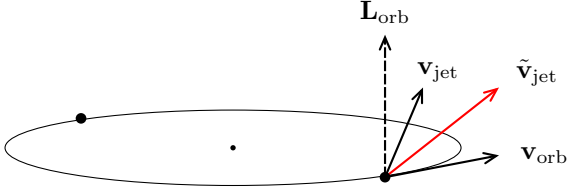

In this paper, we study the morphology of jets from binary BHs whose
dynamics is dominated by GW emission.  Jets in binary BHs have been
studied in a number of previous works \citep{1980ApJ...238L.129S,
  1981ApJ...246L..65I, 1981ApJ...246L.141H, 1982ApJ...258L..63G,
  1982ApJ...253L...1G, 1982ApJ...262..478G}.  It has been shown that
the orbital motion of binary results in a precession of the jet,
resulting in a helical or ``wiggly'' jet morphology.  Such models of
jet precession have been used to understand the observed jet
morphology in several AGN.  For example, recently
\citet{2014MNRAS.445.1370K} fit a jet precession model to the observed
jet in the quasar S5 1928$+$738 using 18 yr of very long baseline
interferometric (VLBI) data.  This object was first studied using BH
binary models by \citet{1993ApJ...409..130R}, but
\citet{2014MNRAS.445.1370K} also included effects of BH spin.  Other
objects studied in this manner are PG 1302-102
\citep{2015MNRAS.454.1290K, 2015Natur.518...74G}, BL Lacertae
\citep{2003MNRAS.341..405S, 2013MNRAS.428..280C}, NGC 4151
\citep{2012ApJ...759..118B}, OJ 287 \citep{2012MNRAS.421.1861V}, S5
1803$+$784 \citep{2008A&A...483..125R}, 3C 345
\citep{2005A&A...431..831L}, 3C 210 \citep{2004MNRAS.349.1218C}, PKS
420-014 \citep{2001A&A...374..784B}, 3C 273
\citep{2000A&A...360...57R}, and Mrk 501 \citep{1999A&A...347...30V}.
\citet{1982ApJ...262..478G} also studied several AGN using their model
of jet precession.

The shortest binary separations inferred from these studies are of the
order of $10^{-2}$ pc.  At smaller separations of around $10^{-3}$ pc,
the orbital velocity of the BHs typically becomes relativistic
($v_\mathrm{orb}\sim 10^4$ km$/$s or $\beta\sim 0.01$).  It was
thought that at these small separations, the binary evolves in a
vacuum, because the inspiral time scale of the binary is likely
shorter than the viscous time scale of any circumbinary gas disk.  As
a result, gas fails to catch up with the BHs, making any accretion
activity impossible in this stage of the binary's evolution
\citep{2005ApJ...622L..93M, 2010PhRvD..81b4019S, 2010ApJ...714..404T}.
However recent viscous hydrodynamical simulations do not find this
decoupling between the BHs and the circumbinary disk.  In these
simulations, the binary's member BHs can continue to accrete gas from
the circumbinary accretion disk all the way until coalescence
\citep{2012PhRvL.109v1102F, 2013MNRAS.436.2997D, 2014PhRvD..89f4060G,
  2014PhRvD..90j4030G, 2015MNRAS.446L..36F, 2015ApJ...807..131S}.
This opens up the possibility of jet production by one of the binary
members during the binary's final phase of evolution before
coalescence, i.e., when the binary's dynamics is dominated by GW
emission.  In this paper we study the morphology of such jets.

All of the above models of jet precession in binary BHs consider
stable binaries, in which the time scale of evolution of the binary
separation is much larger than the orbital period of the binary.  For
such binaries, the helical morphology of the jet is does not change
significantly between precession cycles.  However, we will see below
that for a binary BH with separation $a$, the orbital period is
proportional to $a^{1.5}$, while the time scale of evolution of the
binary separation is proportional to $a^{4}$.  Thus, as the binary
shrinks, the latter time scale decreases much more rapidly than the
former, and for small enough separations, the two time scales can be
comparable.  In this paper, we study jets in this previously ignored
regime.  We show that at this stage of the binary's evolution, the jet
morphology is more complicated: the jet now has a peculiar helical
chirping morphology.  We further argue that in this case the jet
morphology can be used to infer the binary parameters and show how
such features can be distinguished from the usual helical jet
morphology.

\section{Binary parameters}

We study a BH binary formed following a gas-rich merger of two
galaxies as a result of the processes described above.  We consider
black holes with masses $M_1$ and $M_2$ in a circular orbit of radius
$a$ around each other.  Let $M=M_1+M_2$.  For simplicity, we consider
an equal-mass binary ($M_1=M_2$) on a circular orbit. The orbital
speed of each black hole is given by 
\begin{equation}
  v_\mathrm{orb} = \frac{1}{2}\left(\frac{GM}{a}\right)^{1/2}\!\!\!\! = 5.8\times 10^3\, \mathrm{km~s}^{-1}M_8^{1/2}a_{16}^{-1/2},
  \label{eqn:vorb}
\end{equation}
where $a_{16}\equiv (a/10^{16}$ cm$)$ and $M_8\equiv (M/10^8$
M$_\odot)$ are the binary separation and mass,
respectively.\footnote{Note that $10^{16} \mathrm{cm}\approx 3.4\times
  10^{-3} \mathrm{pc}$.  The Schwarzchild radius of a $10^8$~M$_\odot$
  black hole is $9.6\times 10^{-6}$~pc, or $\sim 3\times 10^{13}$~cm.}
The orbital period is
\begin{equation}
  P = 2\pi\left(\frac{GM}{a^3}\right)^{-1/2}\!\!\!\! = 1.72\, \mathrm{yr}\, a_{16}^{3/2} M_8^{-1/2}.
  \label{eqn:period}
\end{equation}

In a gas-rich galaxy merger, the BH binary loses its angular momentum
by torquing the surrounding disk through spiral arms and expelling gas
out of a region twice as large as the binary orbit
\citep{1994ApJ...421..651A, 2005ApJ...622L..93M, 2007PASJ...59..427H,
  2008ApJ...672...83M, 2009MNRAS.393.1423C}.  The expelled gas carries
a specific angular momentum of $\sim va$.  Further angular momentum
dissipation of the binary can happen due to its interaction with fresh
gas that enters the hollowed out region as a result of tidal torques
\citep{2008ApJ...672...83M, 2012ApJ...755...51N, 2012MNRAS.427.2680K,
  2012A&A...545A.127R}.  Therefore, the coalescence time of the binary
is inversely proportional to the rate at which fresh gas re-enters the
binary region.  But only a fraction of gas that enters the central
hollow region accretes onto the black hole and fuels quasar activity.
Therefore, we express $\dot M$ in Eddington units, $\dot{\mathcal
  M}\equiv\dot M /\dot M_E$, where $\dot M_E$ is the accretion rate
required to power the limiting Eddington luminosity with a radiative
efficiency of 10\%, $\dot M_E=2.3 M_\odot\, \mathrm{yr}^{-1}M_8$.  We
therefore have \citep{2010PhRvD..81d7503L, 2015MNRAS.448.3603D}
\begin{equation}
  \frac{a}{\dot a} = t_\mathrm{gas}\approx\frac{J}{\dot Mva}=1.1\times 10^7\,
  \mathrm{yr}\,\dot{\mathcal M}^{-1},
\end{equation}
where $J=\mu va$ is the binary's angular momentum and $\mu=M/2$ is the
binary's reduced mass.  This time scale is greater by two orders of
magnitude than the time scale for coalescence due to angular momentum
loss by GW emission, which for circular orbits is given by
\citep{1964PhRv..136.1224P}
\begin{equation}
  \frac{a}{\dot a} = t_\mathrm{GW} = \frac{5}{256}\frac{c^5a^4}{G^3M^2\mu}=2.53\times 10^5\,\mathrm{yr}\,\frac{a_{16}^4}{M_8^3}.
  \label{eqn:shrink}
\end{equation}
By setting $t_\mathrm{gas}=t_\mathrm{GW}$, we can get the orbital
parameters for the binary when GWs take over.  The orbital speed at
this stage of the binary's evolution is given by
\begin{equation}
  v_\mathrm{orb} = 3.6\times 10^3\, \mathrm{km~s}^{-1}(\dot{\mathcal{M}}M_8)^{1/8},
  \label{eqn:v}
\end{equation}
The separation between the two black holes at this instant is given by
\begin{equation}
  a = 8.3\times 10^{-3}\, \mathrm{pc}\, M_8^{3/4}\dot{\mathcal{M}}^{-1/4}.
  \label{eqn:a}
\end{equation}
Note that the separation is dependent on the gas accretion rate.  

\begin{figure}
  \begin{center}
    \begin{tikzpicture}
      \draw [-angle 45,dashed,thick,dash pattern = on 7pt off 2pt] (0,0) -- (3.0,0);
      \draw [-angle 45,dashed,thick,dash pattern = on 7pt off 2pt] (0,0) -- (0,4.0);
      \draw [-angle 45,thick] (0,0) -- (2.0,4.0);
      \draw [-angle 45,thick,red] (0,0) -- (5.0,4.0);
      \draw (1.0,2.0) arc (63:90:2.2);
      \draw (2.0,1.6) arc (38.6:63:2.56);
      \draw (0.5,2.5) node {\large $\chi$};
      \draw (1.90,2.3) node {\large $\psi$};
      \draw [dashed,thick,dash pattern = on 7pt off 2pt] (0,0) -- (3.0,6.0);
      \draw (0.0,4.5) node {\large $\mathbf{L}_\mathrm{orb}$};
      \draw (3.5,0.0) node {\large $\mathbf{v}_\mathrm{orb}$};
      \draw (1.0,3.0) node {\large $\mathbf{v}_\mathrm{jet}$};
      \draw (5.3,4.2) node {\large \textcolor{red}{$\mathbf{\tilde v}_\mathrm{jet}$}};
    \end{tikzpicture}
  \end{center}
  \caption{The orbital velocity $\mathbf{v}_\mathrm{orb}$
    introduces the jet precession angle $\psi$.  Without the orbital
    motion, $\psi=0$, and the jet traces a cylindrical surface.  With
    the orbital motion, the net red velocity vector oscillates around
    the solid black vector and produces a conical surface with $\psi$
    as its half-opening angle.  Note that $v_\mathrm{orb}$ is
    exaggerated here.  In practice, $v_\mathrm{orb}\ll
    v_\mathrm{jet}\sim c$.  The figure indicates that
    $\sin{\psi}\sim\tan{\psi}=v_\mathrm{orb}\cos{\chi}/v_\mathrm{jet}$.}
  \label{fig:jet}
\end{figure}
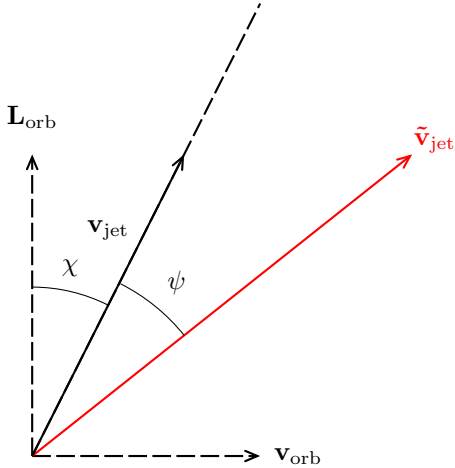

In the GW dominated phase of binary's evolution, recent simulations
show that the BHs can continue to accrete gas from the circumbinary
disk all the way until merger \citep{2015MNRAS.446L..36F}.  At this
stage, we consider a ballistic jet emanating from one of binary's
member BHs at speed $\beta=v_\mathrm{jet}/c$ at an angle $\chi$ to the
direction of the orbital angular momentum.  This geometry is depicted
in Figures~\ref{fig:BH_orbit} and \ref{fig:jet}.  As a result of the
binary orbit, the jet will wiggle on a conical surface.  (In the limit
of a small binary velocity, i.e., at large relative separations, the
jet material will move on a cylindrical surface.)  The half-opening
angle of the cone is given by \citep{1993ApJ...409..130R}
\begin{equation}
  \sin\psi = \frac{v_\mathrm{orb}\cos\chi}{v_\mathrm{jet}}.
\end{equation}
(Note that this relation is valid in the $v_\mathrm{orb}\ll
v_\mathrm{jet}\sim c$. Also, we are ignoring any relativistic
corrections to this equation.)

When the binary separation reaches the point where gravitational
wave-induced inspiral begins, the binary has a velocity given by
equation~(\ref{eqn:v}), and a separation given by
equation~(\ref{eqn:a}).  From equation (\ref{eqn:vorb}), the orbital
velocity of each BH evolves as
\begin{equation}
  \frac{\dot v_\mathrm{orb}}{v_\mathrm{orb}} = -\frac{1}{2}\frac{\dot a}{a}.
\end{equation}
which gives the evolution of the opening angle of the cone as 
\begin{equation}
  \dot\psi = -\frac{1}{2\cot\psi}\frac{\dot a}{a}.
  \label{eqn:widen}
\end{equation}
This can be solved analytically to get
\begin{equation}
  \sin\psi = \sin\psi_0\left(\frac{a_0}{a}\right)^{1/2},
\end{equation}
where $\psi_0$ and $a_0$ are the initial values. Thus, as the binary
continues to shrink the BH orbital velocity grows and the opening
angle of the conical jet increases (cone opens wider).  The rate at
which this happens reflects the binary's dynamical evolution.

\section{Jet precession}

\begin{figure}
  \begin{center}
    \begin{tikzpicture}
      \draw [thick,color=brown] (0,0) -- (-4.0,0);
      \draw [thick,color=brown] (0,0) -- (0,4.0);
      \draw [thick,color=brown] (0,0) -- (2.6,-3.0);
      
      \draw [dashed,thick,dash pattern = on 7pt off 2pt,color=red] (0.05,0.0) -- (0.05,4.0);
      \draw [dashed,thick,dash pattern = on 7pt off 2pt,color=red] (0,0.0) -- (3.46,-2.0);
      \draw [dashed,thick,dash pattern = on 7pt off 2pt,color=red] (0,0.0) -- (-3.87,-1.0);

      \draw [thick,rotate around={-75:(-2.95,-0.76)}] (-2.95,-0.76) ellipse (0.55 and 0.2);
      \draw [thick,rotate around={12:(3.04,0.24)}] (3.04,0.238) arc (-90:90:0.2 and 0.55);
      \draw [thick,color=gray,rotate around={12:(3.04,0.24)},dashed](3.04,0.238) arc (270:90:0.2 and 0.55);
      
      \draw [thick] (-2.82,-1.3) -- (2.82,1.3);
      \draw [thick] (-3.09,-0.24) -- (3.09,0.24);

      \draw (0.1,4.3) node {\large \textcolor{brown}{$y$},\textcolor{red}{$\,y^\prime$}};
      \draw (-4.2,0) node {\textcolor{brown}{\large $z$}};
      \draw (2.75,-3.17) node {\textcolor{brown}{\large $x$}};
      \draw (3.1,-3.57) node {\large [line of sight]};
      \draw (-4.1,-1.059) node {\large \textcolor{red}{$z^\prime$}};
      \draw (3.7,-2.1) node {\textcolor{red}{\large $x^\prime$}};

      \draw [thick,color=brown] (-1.0,-0.258) arc (230:283:1.8);
      \draw (-0.5,-1.0) node {\textcolor{brown}{\large $i$}};

      \draw (-2.0,-0.92) arc (204.75:194.48:2.2015);
      \draw (-2.0,-1.3) node {\large $\psi$};

    \end{tikzpicture}
  \end{center}
  \caption{Geometry of the jet \citep{1982ApJ...262..478G}.  The
    observer's sky is in the $yz$ plane, so that the line of sight is
    along the $x$-axis.  The jet precesses around the $z^\prime$ axis
    with half-opening angle $\psi$. The $y$ and $y^\prime$ axes
    overlap, and $z$, $x$, and $z^\prime$ axes are co-planer.  The
    angle between the $z^\prime$ axis and the line of sight is $i$.
    This geometry is valid in the limit $a\ll z^\prime\psi$.}
  \label{fig:jetgeom}
\end{figure}

\begin{figure*}
\begin{center}
  \begin{tabular}{cc}
    \includegraphics*[scale=0.45]{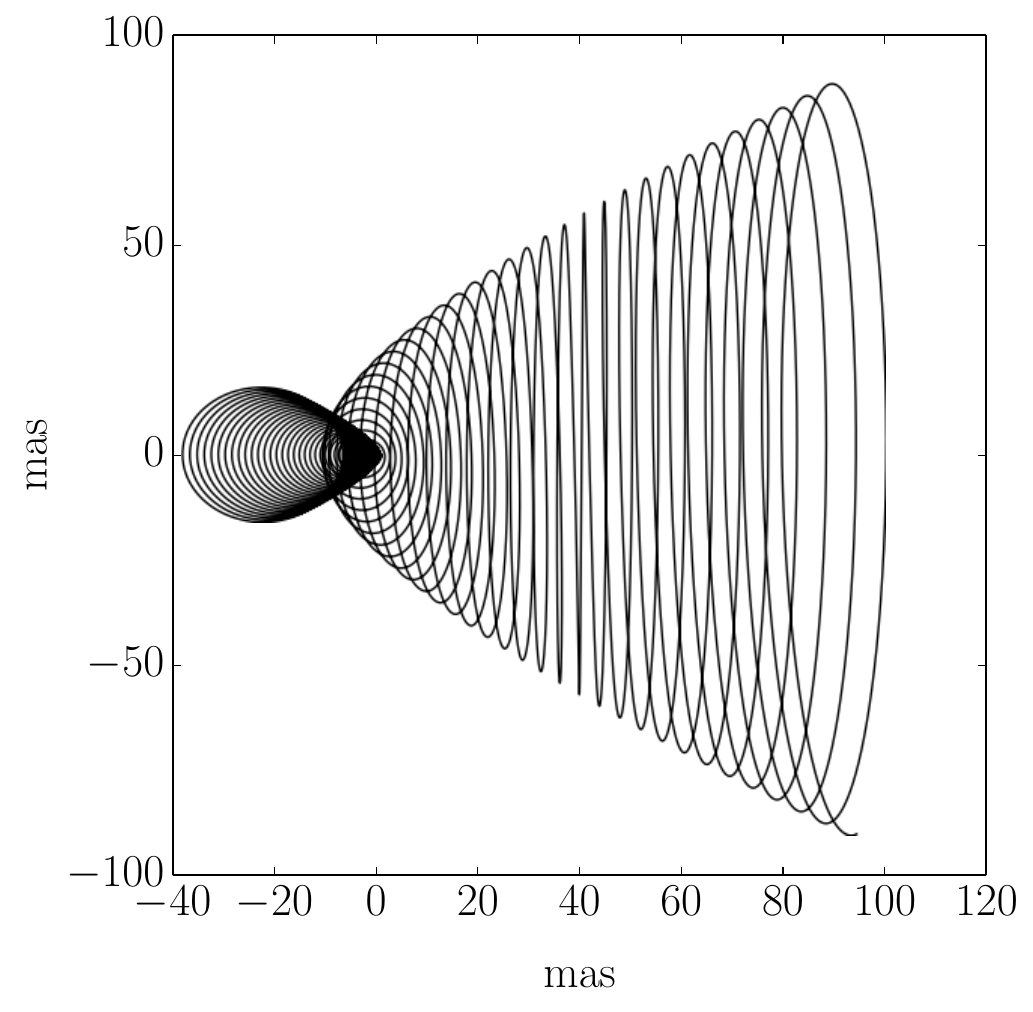} &
    \includegraphics*[scale=0.45]{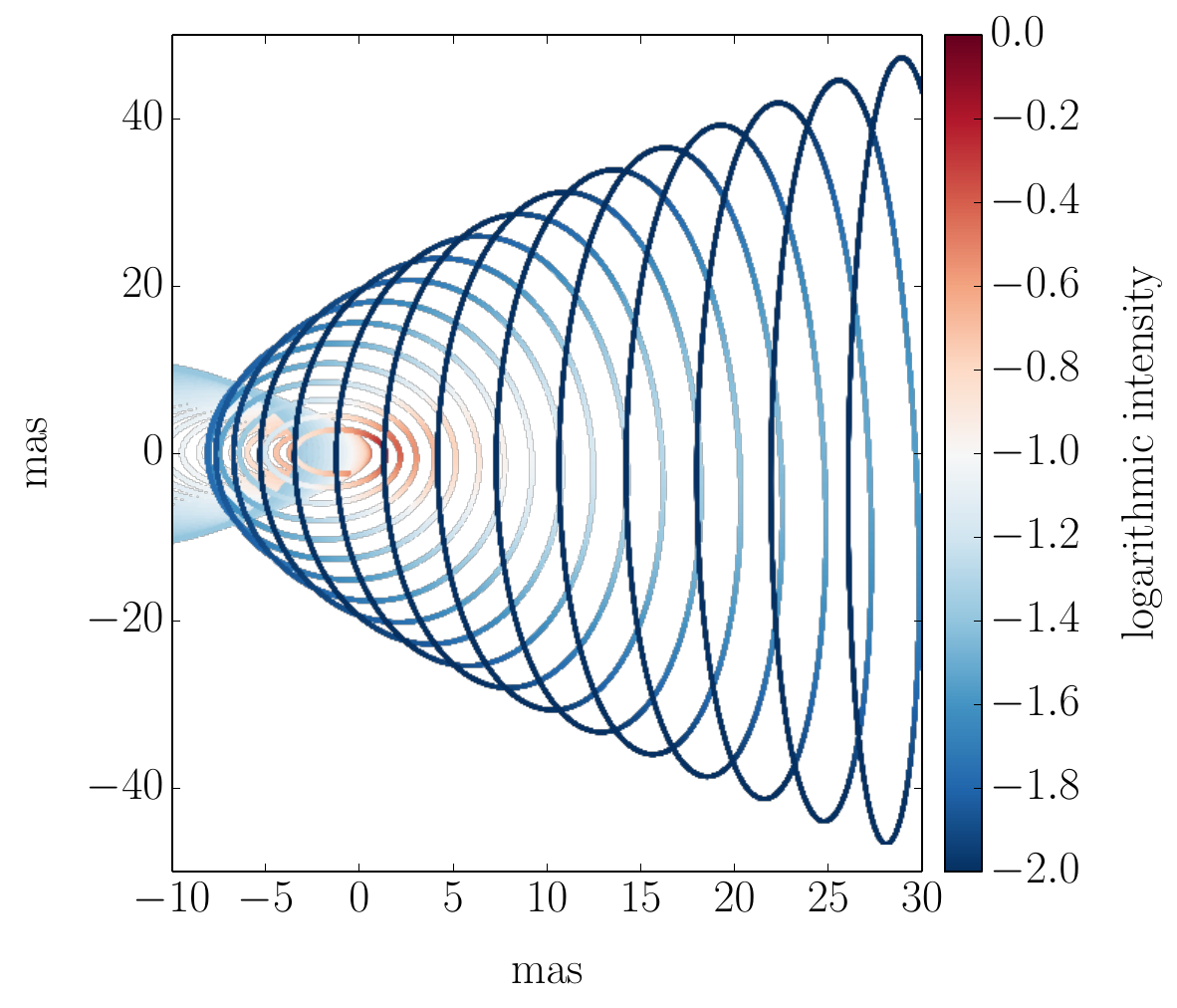}
  \end{tabular}
\end{center}
\caption{The left panel shows forward and backward jets for an
  equal-mass binary with total mass $M=10^{10}$~M$_\odot$.  (This
  figure shows the position of many jet particles observed at a fixed
  time on Earth.)  We have assumed $i=40^\circ$, $\beta=0.9$, and
  angular diameter distance $d = 100$~Mpc.  The jets are twisted
  because of the apparent superluminal motion.  Also evident is the
  opening of the jet and the radio jet chirp.  Right panel shows
  normalised brightness of the inner 30~mas region of the forward
  jet. (This panel shows the intensity of many particles observed at a
  fixed time on Earth.)}
\label{fig:jet_mdot1.0}
\end{figure*}

We now ask if it would be possible to observe the opening of the jet
morphology.  For this purpose, it is necessary to map the above
calculations to the observer's frame \citep{1982ApJ...262..478G}.  At
a large distance compared to the binary's separation, the jet geometry
is illustrated in Figure \ref{fig:jetgeom}.  Here the source frame of
reference is shown by $x^\prime$, $y^\prime$, $z^\prime$ coordinate
systems.  The $x$, $y$, $z$ coordinate system represents the
observer's frame such that the $yz$ is the plane of the sky.  Further,
the jet is oriented such that the $y^\prime$ axis coincides with the
$y$ axis, the original direction of jet emission is at an angle $i$
relative to the line of sight ($x$-axis).  For simplicity, we also
assume that the $z^\prime$ axis is in the $xz$ plane.  Then the jet
will precess on a cone with a half-opening angle $\psi$ and an angular
velocity of precession $\Omega$.  In the source frame of reference,
the equation of the jet is then given by \citep{1981ApJ...246L.141H}
\begin{eqnarray}
  &v_x=&s_\mathrm{jet}\beta c\left[\sin\psi\sin i\cos\Omega t+\cos\psi\cos i\right],\\
  &v_y=&s_\mathrm{jet}\beta c\sin\psi\sin\Omega t,~\,\mathrm{and}\\
  &v_z=&s_\mathrm{jet}\beta c\left[\cos\psi\sin i-\sin\psi\cos i\cos\Omega t\right]. 
\end{eqnarray}
Here $s_\mathrm{jet}=\pm 1$ for forward and backward jets,
respectively.  The coordinates of a jet particle in the sky is then
given by $z=v_zt$ and $y=v_yt$.  However, because of superluminal
motion, the observer will see this on the sky as
\begin{equation}
  z=v_zt/(1-v_x/c)
\end{equation}
and
\begin{equation}
  y=v_yt/(1-v_x/c)
\end{equation}
respectively.  In angular coordinates, the motion of the jet on the
sky is given by $\phi_z=z/d$ and $\phi_y=y/d$ where $d$ is the angular
diameter distance of the binary.  Because of superluminal motion, the
forward jet is stretched while the backward jet is compressed.

Finally, we assume that the emission from each jet is optically thin
at the frequency of observation with a spectrum in the rest frame of
the particle given by $P(\nu)\propto \nu^{-\alpha}$
\citep{1982ApJ...262..478G}.  The observed flux density $S(\nu)$ in
erg cm$^{-2}$ Hz$^{-1}$ is given by
\begin{equation}
  S(\nu)=S_r(\nu)D^{3+\alpha},
\end{equation}
where $D$ is the Doppler shift parameter
$[\gamma(1-\beta\cos\phi)]^{-1}$ and $S_r$ is the rest frame flux
density.  Further, to consider the effects of jet evolution we assume
that $S_r$ decreases with time according to a simple power law
\begin{equation}
  S_r\propto t_\mathrm{proper}^{-\delta},
\end{equation}
where $t_\mathrm{proper}$ is the proper time in the frame of the jet.
We assume $\delta=1$ and $\alpha=1$, hereafter
\citep{1982ApJ...262..478G}.

\section{Jet Morphology}

As the binary evolves, the combination of increasing orbital speed and
the apparent superluminal motion effect results in a jet morphology
with an increasing opening angle and increasingly rapid periodic
variation, together with a twisted shape.
Figure~\ref{fig:jet_mdot1.0} shows these three features for an
equal-mass binary with total mass $M=10^{10}$~M$_\odot$.  For
concreteness, we have assumed $i=40^\circ$ and $\beta=0.9$.  The
angular diameter distance to the binary is assumed to be $d = 100$~Mpc
(which corresponds to redshift $z\sim 0.025$).  The forward jet
stretches on the sky while the backward jet is compressed due to the
apparent superluminal motion.  The right panel of
Figure~\ref{fig:jet_mdot1.0} zooms into the inner region of the
forward jet and shows that the binary creates wiggles in the jet
morphology of the order of a milliarcsecond.  The duration of the
binary evolution captured by Figure~\ref{fig:jet_mdot1.0} is 100 yr.
Although the system can be rescaled to different masses, times, and
separation, the initial separation in Figure~\ref{fig:jet_mdot1.0} is
$\sim 10^{-3}$~pc, when the binary enters it GW-dominated phase.

The increasing orbital speed of the binary leads to an increase in the
jet's precession angular velocity.  This increases the helical winding
of the jet such that the jet is wound up increasingly tightly closer
to the source.  This gives rise to a ``chirping'' morphology as seen
in Figure~\ref{fig:jet_mdot1.0}.  The chirp is easier to resolve in
jets that are closer to the line of sight, where superluminal motion
can magnify the forward jet.  A chirping black hole is an unambiguous
signature of a rapidly evolving binary BH system; the chirp is direct
reflection of the orbital velocity evolution of the binary. Together,
the three features of the jet morphology in
Figure~\ref{fig:jet_mdot1.0} are:
\begin{itemize}
\item Increase in the helical winding of the jet towards the central source (``chirpyness''),
\item Widening of the opening angle towards the central source, and 
\item An apparant variation in the jet precession axis towards the central source (``twist''). 
\end{itemize}
These directly probe the dynamics of the binary in its GW phase.

\subsection{Can chirping jets be observed?}

Observations of such a milliarcsecond scale geometry with the three
features described above can thus flag a BH binary nearing the end of
its merger.  We now consider the question of whether such morphologies
will be measurable in real observations.

As discussed in Section~\ref{sec:intro}, several helical jets have
been observed thanks to VLBI.  A resolution of few tens of microarc
seconds can be achieved using the Very Long Baseline Array (VLBA) with
sensitivity of up to 30 $\mu$Jy$/$beam, as demonstrated by the MOJAVE
survey \citep{2009AJ....138.1874L}.  VLA has angular resolution of few
tens of milliarcsecconds with a sensitivity of about 100
$\mu$Jy$/$beam.  The European VLBI network (EVN) which can resolve up
to 0.13 milliarcseconds and can detect surface brightness down to 19
$\mu$Jy$/$beam.  The Very Large Array (VLA) has an angular resolution
of few tens of milliarcsecconds with a sensitivity of about 100
$\mu$Jy$/$beam.  The future Square Kilometer Array (SKA) on its own
will have 2 milliarcsecond resolution at around 10 GHz, but SKA will
be an order of magnitude more sensitive than current instruments with
a sensitivity of about 50 nJy$/$beam.  Thus angular resolution of tens
to hundreds of microarc seconds and flux sensitivity of a few
$\mu$Jy$/$beam is achievable with current VLBI instruments.

\begin{figure*}
\begin{center}
  \begin{tabular}{cc}
    \includegraphics*[width=\columnwidth]{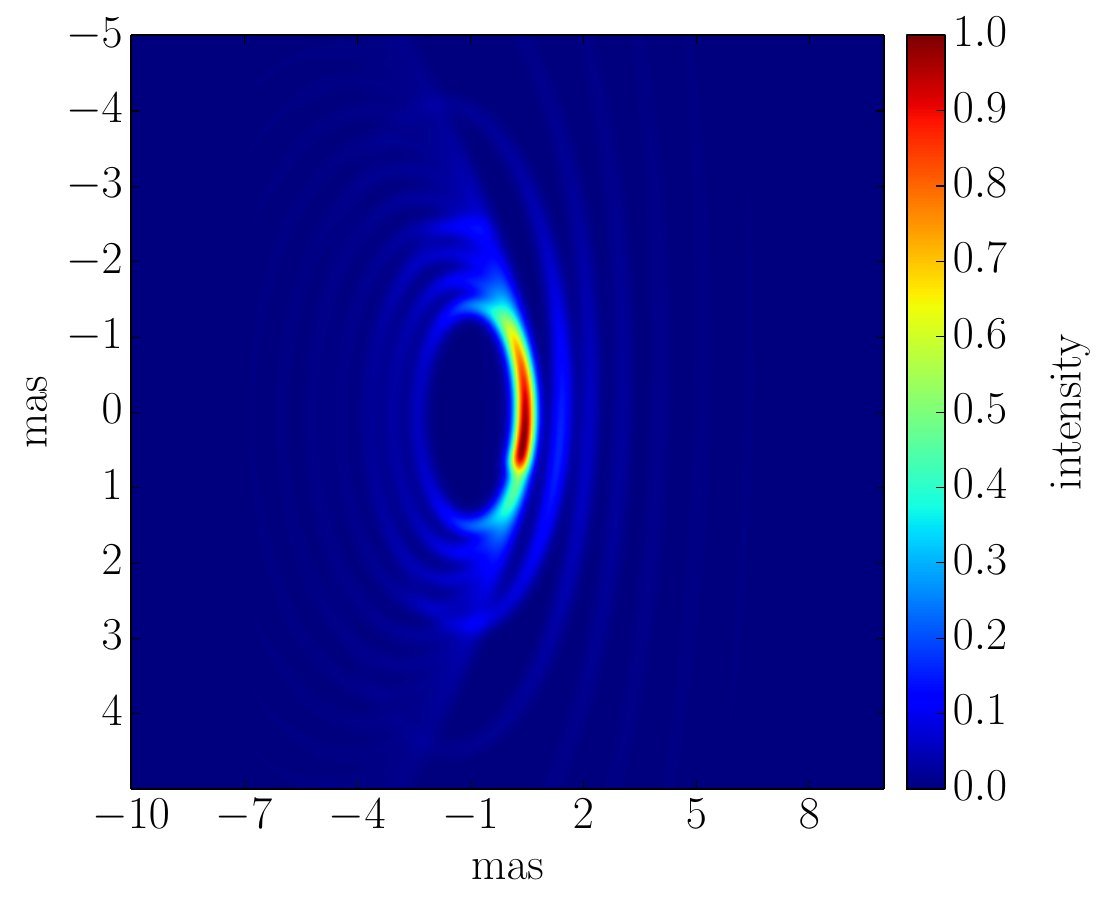} &
    \includegraphics*[width=\columnwidth]{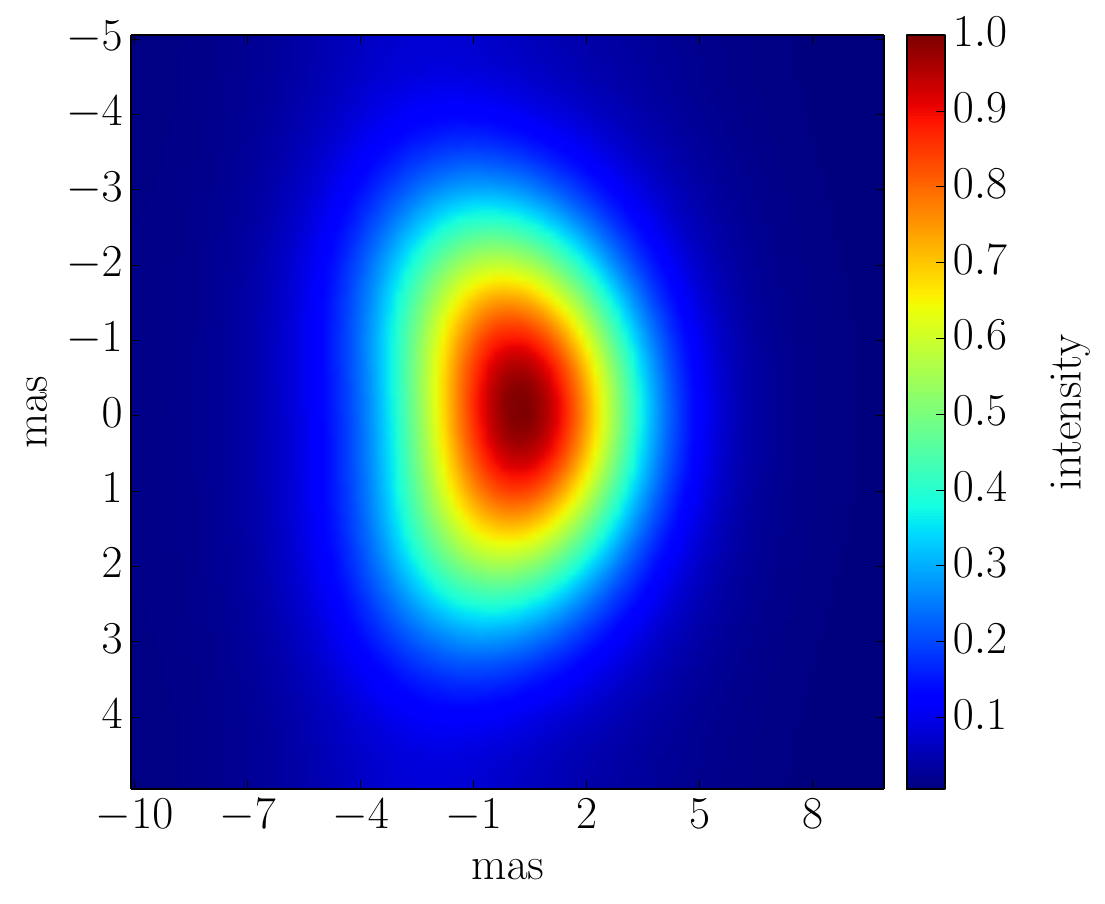}
  \end{tabular}
\end{center}
\caption{Central 100 square mas of the jet in
  Figure~\ref{fig:jet_mdot1.0} smoothed at 200 $\mu$as (left panel)
  and 2 mas (right panel) respectively.  The intensity is normalised
  in arbitrary units.}
\label{fig:smooth_jet}
\end{figure*}

The brighter part of the jet (intensity within factor of 10 from the
maximum) in Figure~\ref{fig:jet_mdot1.0} (right panel) has separations
of $\sim 1$ mas.  Figure~\ref{fig:smooth_jet} shows inner 100 square
mas of the jet in Figure~\ref{fig:jet_mdot1.0} smoothed at 200 $\mu$as
and 2 mas respectively.  We find that at 200 $\mu$as resolution the
chirp is clearly resolved.  At a lower resolution of 2 mas, the chirp
is no longer resolved but widening of the jet opening angle can
possibly be inferred.  (In the next section, we discuss an algorithm
to fit the jet model to observational data.)  Note that other
geometric effects can influence these conclusions.  If it is aligned
closer to the line of sight, the forward jet is magnified due to
apparant superluminal motion.  This will make it possible to resolve
jets of objects that are farther away than the example in
Figure~\ref{fig:jet_mdot1.0}.  Binaries with higher masses have large
orbital periods that will result in a helix with wider separation
between its cycles.  These objects would therefore be easier to
detect.  We have not modeled the absolute surface brightness of jets,
only relative variation in the jet brightness (as shown in
Figure~\ref{fig:jet_mdot1.0}), but due to the $(1+z)^{-4}$ dimming, it
will be exceedingly difficult to detect such jets at higher redshifts.

We note here that there is at least one object in the literature that
shows a jet morphology similar to Figure~\ref{fig:jet_mdot1.0}.  The
quasar 4C 18.68 described by \citet{1982ApJ...253L...1G} shows a
helical jet with an apparant opening angle that increases towards the
central source from 11$^\circ$ to 45$^\circ$ in just two cycles
\citep{1982ApJ...262..478G}.  The observing beamwidth of about 400 mas
makes it difficult to resolve a chirp in this source, but it is a
prototype of the morphology that we are discussing here.
\citet{2015MNRAS.453.1562G} have reported that 111 candidates show
periodic variability in brightness in the Catalina Real-time Transient
Survey, indicating presence of binary BHs.  An analysis of these
candidates suggests the presence of circumbinary gas at small orbital
distances \citep{2015MNRAS.453.1562G}.  This bodes well for the
likelihood of the presence of chirpy jets.  In low resolution wide
field surveys, detecting modulations in the light curves would be an
easier diagnostic of a binary BH \citep{2011ApJ...743..136O,
  2011ApJ...734L..37K}.

\subsection{Deriving binary parameters from observations of chirping jets}
\label{sec:geometric_model}

From observations of chirping jets, we can infer the dynamical
parameters of the binary BH by first fitting a purely geometric model
to the observed jet morphology.  To create a model of this kind, we
generalise the simple helical model considered by
\citet{2014MNRAS.445.1370K} and \citet{2001NewA....6...61N} to write
\begin{align}
  z &= x^{-\alpha} \frac{B}{2\pi} u \cos{u}, \label{eqn:z}\\
  y &= x^{-\alpha} \frac{B}{2\pi} u \sin{u}, \\
  x &= \frac{A}{2\pi} u \left(\frac{u}{2\pi}\right)^\beta.
\end{align}
Here $u$ is the azimuthal angle around the jet axis, $x$ and $y$ axes
are in the plane of the sky, and the $z$ axis points to the observer.
The $x^{-\alpha}$ term in the first two equations captures the
widening opening angle of the jet, while the $(u/2\pi)^{\beta}$ term
in the last equation parameterises the chirp.  This model can be
fitted to an observed jet on the sky by using two rotation matrices,
$\mathcal{R}_\lambda$ and $\mathcal{R}_i$, which rotate the model in
the above equations to match with the jet precession axis and the jet
symmetry axis.  These matrices are given by

\begin{flalign}
\mathrm{R}_{i} \!=\!
\begin{pmatrix}
\cos i & 0 & \sin i \\ 
0 & 1 & 0 \\ 
-\sin i & 0 & \cos i
\end{pmatrix}
\label{riota}
\end{flalign}
and
\begin{flalign}
\mathcal{R}_{\lambda}\!=\!
\begin{pmatrix}
\cos \lambda{_0} & -\sin \lambda{_0} & 0 \\ 
\sin \lambda{_0} & \cos \lambda{_0} & 0 \\ 
0 & 0 & 1%
\end{pmatrix}.
\label{rlambda}
\end{flalign}
Further, we can parameterise the ``twist'' in the jet by requiring
that $i$ and $\lambda$ vary with $x$.  Without loss of generality,
we take this variation in $i$ and parameterise it as
\begin{equation}
  i = \gamma x + i_0.
  \label{eqn:i}
\end{equation}
We thus arrive at a model with 7 parameters ($\alpha$, $\beta$,
$\gamma$, $A$, $B$, $i_0$, $\lambda_0$), which can be fitted to any
observed jet, such as that in Figure~\ref{fig:jet_mdot1.0}.  Note that
setting $\alpha$ and $\beta$ equal to $0$ yields the usual helical jet
morphology, without the features discussed in this paper.  Thus, a
non-zero value of these two parameters distinguishes chirpy jets from
non-chirpy ones.

\begin{figure*}
\begin{center}
  \includegraphics*[width=\textwidth]{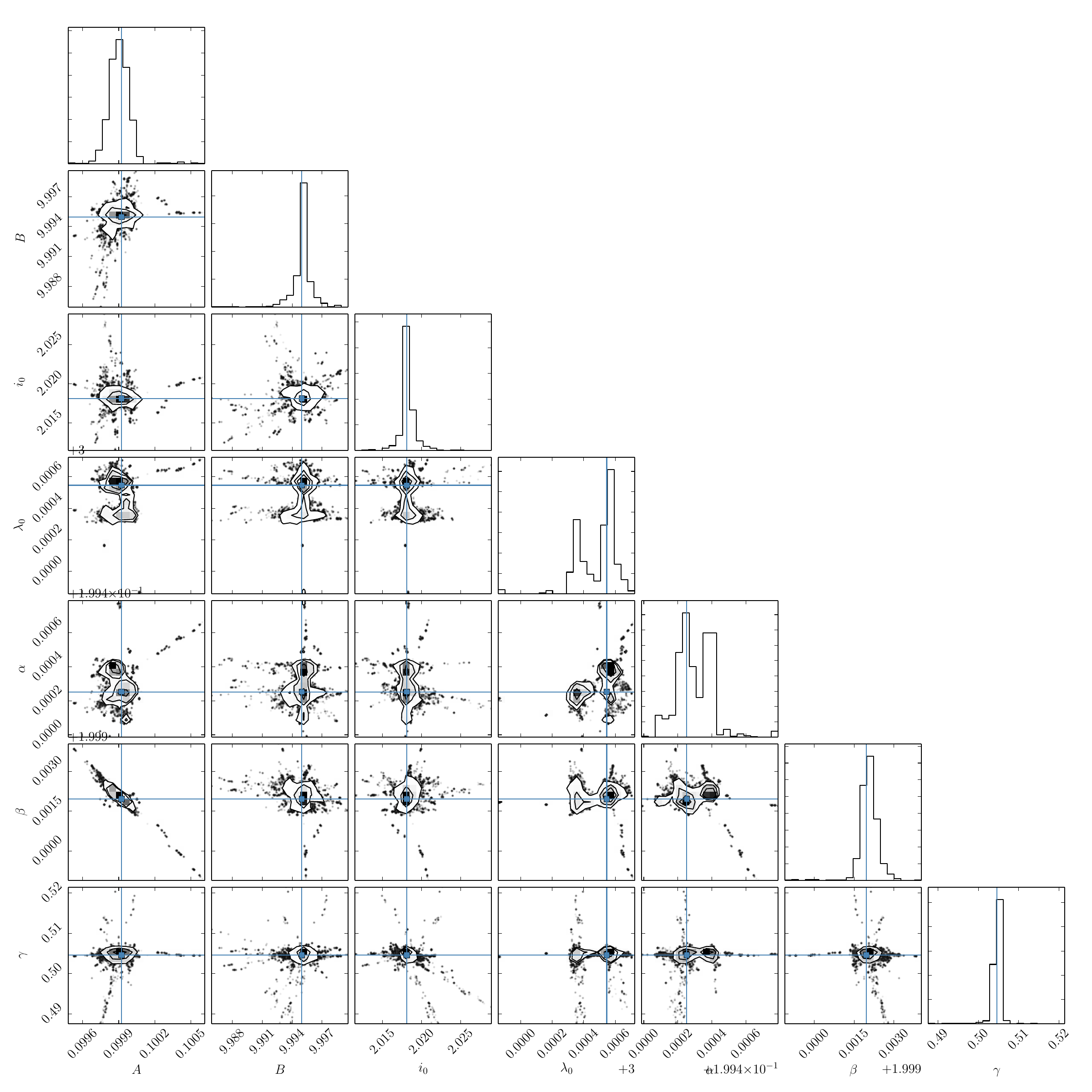}
\end{center}
\caption{One and two dimensional projections of posterior probability
  distributions of the 7-parameter geometric model described in
  Section~\ref{sec:geometric_model} to mock data.  Wide, uniform
  priors were chosen for all parameters.  The two-dimensional
  projections also show that there are no parameter degeneracies in
  the geometric model.  As described in the text, these parameter
  values can then be used to derive various physical parameters
  describing the black hole binary.}
\label{fig:triangle}
\end{figure*}

Figure \ref{fig:triangle} shows how the geometric model of this
section works in practice, by fitting it to mock data using Markov
Chain Monte Carlo (MCMC).  We used the \texttt{emcee} package for this
purpose \citep{2013PASP..125..306F}.  Mock data are created by using
Equations~(\ref{eqn:z})--(\ref{eqn:i}) and then smoothing the
resultant jet morphology by $\sim$ 2 mas.  Figure \ref{fig:triangle}
shows posterior probability distributions of the parameters of our
7-parameter model when it is fit to these mock data.  Wide, uniform
priors were chosen for all parameters.  The posteriors demonstrate
that there is no degeneracy in the geometric model parameters.  In
fact the narrow ranges of values for which the posteriors are non-zero
for each parameter show that the model fit the mock data very well.

The best fit geometric parameters can then be used to derive the
dynamical parameters of the BH binary in the following way.  The
parameters $\alpha$ and $B$ quantify the widening of the opening angle
$\psi$.  As each cycle of the observed helical jet corresponds to a
period $P(t)$ of the binary, we can then write $d\psi/dP$, which
yields a constraint on the quantity $a^4/M^3$ as a function of $P$ via
Equations (\ref{eqn:period}), (\ref{eqn:shrink}) and
(\ref{eqn:widen}).  The parameters $A$ and $\beta$, which quantify the
chirp, can be used to directly infer the evolution of the binary
separation as $da/dP$, which also constraints the quantity $a^4/M^3$
via Equation~(\ref{eqn:shrink}). These two constraints can be used to
determine $a(P)$ and $M$ separately.  Finally, the parameters
$i_0$, $\lambda_0$, and $\gamma$ quantify the twist in the jet,
which can potentially be used to constrain the quantity
$\dot{\mathcal{M}}$ by using Equation~(\ref{eqn:v}).

Further parameters can be introduced in the dynamical model by
considering unequal-mass binaries, eccentric orbits, and
Post-Newtonian corrections, but we not develop this further in this
paper.  We make the code for implementing the model presented in this
paper and for fitting it to observed jets using MCMC analyses
available at \url{https://github.com/gkulkarni/JetMorphology} under
the MIT open source software license.  This code can be used to
explore parameter space of the model.

\section{Conclusion}

Using a model of jet precession in a binary BH, we have shown that the
jet produced by a member of a binary that is entering its GW-induced
phase of inspiral exhibits a distinct chirping morphology.  Jets in
these binaries have a biconical morphology.  After the binary enters
the GW phase, the opening angle of cone rapidly increases within a
separation of tens of milliarcseconds at a distance of $\sim 100$~Mpc,
depending on the angle that the jet makes relative to the line of
sight.  In addition, the periodicity of the jet's motion on the
conical surface increases with the orbital period of the binary in
this stage of its evolution.  Finally, due to the apparent
superluminal motion and the widening opening angle, the jet is twisted
into a peculiar morphology.  The jet geometry can be used to infer the
set of dynamical and geometrical parameters.

Milliarcsecond features can be resolved using long baseline radio
interferometry \citep{2001ChJAA...1..236Q}.  At higher redshifts, jets
are fainter and the wiggles become more difficult to resolve.
However, it might be possible to resolve binaries with very unequal
mass ratios, eccentric orbits, or suitable inclinations.  The future
Square Kilometer Array (SKA) should make microarcsecond resolution
possible, which could increase the detection rate of these objects
\citep{2015aska.confE.143P}.

The jet morphology acts as an electromagnetic counterpart for chirping
gravitational wave sources.  BH binaries are expected to be the
brightest GW sources in the universe.  Current ground based GW
detectors such as LIGO are sensitive to intermediate-mass BH binaries
at high redshift, which would be difficult to detect with our method.
However, future detectors such as eLISA are sensitive to GW emission
from $10^7$--$10^9$ M$_\odot$ BHs \citep{2013CQGra..30x4009S}.  Pulsar
Timing Arrays are sensitive to even higher binary masses.  Monitoring
of jet morphologies for the features outlined in this paper will put
a lower bound on the expected abundance of bright GW
sources. Moreover, combined GW and electromagnetic observations of
compact binary mergers should enable detailed studies of astrophysical
processes in the strong-field GR regime \citep{2005ApJ...629...15H}.

\section*{Acknowledgments}

We acknowledge helpful comments from the anonymous referee and useful
discussions with Martin Haehnelt, Enrico Ramirez-Ruiz, and Alberto
Rorai.  This work was supported in part by NSF grant AST-1312034 (for
A.~L.) and by the FP7 ERC Advanced Grant Emergence-320596 (for G.~K.).

\bibliographystyle{mnras}
\bibliography{refs} 

\bsp
\label{lastpage}
\end{document}